\documentclass[pra,twocolumn,floatfix,aps,superscriptaddress,showpacs,amsfonts]{revtex4}
\usepackage{graphicx}
\usepackage{array}
\usepackage{amsthm}
\usepackage{amsmath}
\usepackage{amssymb}

\begin{document}
\title{One-way deficit and quantum phase transitions in $XX$ Model}

\author{Yao-Kun Wang}
\email{wangyaokun521@126.com}
\affiliation{College of Mathematics,  Tonghua Normal University,
 Tonghua, Jilin 134001, P. R. China}
 \affiliation{Institute of Physics, Chinese Academy of Sciences, Beijing 100190, P. R. China}




\begin{abstract}
 Quantum correlations including entanglement and quantum discord has drawn much attention in characterizing quantum
phase transitions. Quantum deficit originates in questions regarding work extraction from quantum systems
coupled to a heat bath [Phys. Rev. Lett. \textbf{89}, 180402 (2002)].
It links quantum thermodynamics with quantum correlations and provides a new standpoint for understanding quantum non-locality.
In this paper, we evaluate the one-way deficit of two adjacent spins in the bulk for the XX model. In the thermodynamic limit, the XX model undergoes a first order transition from fully polarized to a critical phase with quasi-long-range order with decrease of quantum parameter. We find that the one-way deficit becomes nonzero
 after the critical point. Therefore, the one-way deficit characterizes
the quantum phase transition in the XX model.
\end{abstract}
\eid{identifier}
\maketitle

\section{Introduction}
The recent development in quantum information theory \cite{key-1} has provided much insight into quantum phase transitions \cite{key-2}.
Especially, using the quantum correlations to research quantum
phase transitions has attracted much attention and has been successful
in characterizing a large number of critical phenomena of great interest.
In this family, entanglement was the first and most outstanding member to detect various of quantum phase transitions, see \cite{key-3,key-4,key-5,key-6,key-7}.
Quantum discord is a measure of the difference between the  mutual information and maximum classical mutual information, is also used to study quantum phase transitions \cite{key-8,key-9}. Another indication of quantumness that is also found its applications for probing quantum phases and quantum phase transition \cite{key-10,key-11,key-12,key-13}.

Besides entanglement and quantum discord, quantum deficit\cite{oppenheim,horodecki,modi2} originates on asking how to use
nonlocal operation to extract work from a correlated system coupled to a heat bath \cite{oppenheim}.
 Oppenheim \emph{et al.} defined the work deficit \cite{oppenheim}
is a measure of the difference between the information of the whole system and the localizable information\cite{horodecki2,ho03}.
Recently, by means of relative entropy, Streltsov \emph{et al. }\cite{Streltsov0,chuan} give the definition of the one-way information deficit which is also called one-way deficit, which uncovers an important role of quantum deficit as a resource for the distribution of entanglement.

In this paper, we will endeavor to calculate the one-way deficit of two adjacent spins in the bulk
of the $XX$ model.
In the thermodynamic limit, it undergoes a first order transition from fully polarized
to a critical phase with quasi-long-range order with decrease of quantum parameter $\lambda$.
We find that the one-way deficit becomes nonzero after the critical point. Therefore,
the one-way deficit characterizes the quantum phase transition in the XX model.
That is, we can employ the deficit to detect the quantum phase transition point  in the XX model.

\section{One-way deficit in $XX$ model for $0<\lambda<1$}\label{sec:FOC}

One-way deficit by von Neumann measurement on one side is given by \cite{streltsov}
\begin{eqnarray}
\Delta^{\rightarrow}(\rho^{ab})=\min\limits_{\{\Pi_{k}\}}S(\sum\limits_{k}\Pi_{k}\rho^{ab}\Pi_{k})-S(\rho^{ab}).\label{definition}
\end{eqnarray}

We investigate the Spin-$\frac{1}{2}$ $XX$ model with the Hamiltonian as
\begin{eqnarray}
H_{xx} = -\frac{1}{2}\sum_{i=1}^{N}\left[\sigma_{i}^x\sigma_{i+1}^x+\sigma_{i}^y\sigma_{i+1}^y \right]-\lambda\sum_{i=1}^{N}\sigma_{i}^z,
\end{eqnarray}
where the interaction constant is taken as the energy unit and $\lambda$ represents the strength of the external
magnetic field and $\sigma^{x}, \sigma^{y}, \sigma^{z}$ are the usual Pauli matrices. Its quantum phase transition is like that in
one phase the system is gapped while in the whole region of the other phase the system is critical. The open
boundary conditions are assumed, i.e. $N+1\equiv 0$. The phase diagram is symmetric in respect of $\lambda$ \cite{XXQPT,SonXX}, therefore we only consider the case of positive $\lambda$. For $\lambda >1$ the ground state is polarized. At $\lambda=1$ the system undergoes a first order quantum transition. In the region $\lambda<1$ the system is critical.  This model can be solved analytically \cite{SonXX} using the following Jordan-Wigner and Fourier transformations
\begin{eqnarray}
d_k=\sqrt{\frac{2}{N+1}}\sum_{l=1}^N\sin\big(\frac{\pi kl}{N+1}\big)\prod_{m=1}^{l-1}\sigma_m^z\sigma_l^-,
\end{eqnarray}
which turn the Hamiltonian into the diagonalized form in terms of fermion operator as $H_{xx}=\sum_{k=1}^N\Lambda_k d_k^\dag d_k+N\lambda$ with $\Lambda_k=2\Big[\cos\big(\frac{\pi k}{N+1}\big)-\lambda\Big]$.

In the phase where $0 \le \lambda < 1$, the ground state corresponds to that having fermions occupied at negative  $\Lambda_k$ only. The reduced density matrix for two spins at $l$ and $l+1$ has been obtained in \cite{SonXX} as
\begin{eqnarray}\label{state1}
\rho_{l,l+1}&=&a_{+}|\uparrow \uparrow\rangle \langle\uparrow \uparrow|+a_{-}|\downarrow \downarrow\rangle \langle\downarrow\downarrow|+ b_{+}|\uparrow \downarrow\rangle \langle\uparrow \downarrow|\nonumber\\
& &+b_{-}|\downarrow \uparrow\rangle \langle\downarrow \uparrow| + e(|\uparrow \downarrow\rangle \langle\downarrow \uparrow|+|\downarrow \uparrow\rangle \langle\uparrow \downarrow|)
\end{eqnarray}
with
\begin{eqnarray}
a_{\pm}=\frac{1}{4}[1\pm\langle\sigma_l^z\rangle\pm\langle\sigma_{l+1}^z\rangle+\langle\sigma_{l}^{z}\sigma_{l+1}^{z}\rangle], \nonumber\\ b_{\pm}=\frac{1}{4}[1\pm\langle\sigma_l^z\rangle\mp\langle\sigma_{l+1}^z\rangle-\langle\sigma_{l}^{z}\sigma_{l+1}^{z}\rangle], \nonumber\\
~~~~~~~~~~~~e=\frac{1}{2}\langle\sigma_{l}^{x}\sigma_{l+1}^{x}\rangle.~~~~~~~~~~~~~
\end{eqnarray}
  In the thermodynamic limit it can be shown that for bulk spins we have\cite{jinjun}
\begin{eqnarray}
\langle\sigma_{l}^{z}\sigma_{l+1}^{z}\rangle &=&\big(1-\frac{2\arccos(\lambda)}{\pi} \big)^2 - \frac{4}{\pi^2}(1-\lambda^2), \nonumber\\
\langle\sigma_{l}^{x}\sigma_{l+1}^{x}\rangle &=& - \frac{2}{\pi}\sin(\arccos(\lambda)), \nonumber\\
\langle\sigma_l^z\rangle&=&\langle\sigma_{l+1}^z\rangle=1-\frac{2\arccos(\lambda)}{\pi}.
\end{eqnarray}

Let
\begin{eqnarray}
c=c_1&=&c_2=\langle\sigma_{l}^{x}\sigma_{l+1}^{x}\rangle, \nonumber\\
c_3&=&\langle\sigma_{l}^{z}\sigma_{l+1}^{z}\rangle, \nonumber\\
r&=&s=\langle\sigma_l^z\rangle.
\end{eqnarray}

The density matrix in Eq. (\ref{state1}) is rewritten as
\begin{widetext}
\begin{eqnarray}\label{state3}
\rho_{l,l+1} &=&\frac{1}{4}
\left(
\begin{array}{cccc}
1+2r+c_3
& 0 & 0 & 0 \\
0 & 1-c_3 & 2c & 0 \\
0 & 2c & 1-c_3
& 0 \\
0 & 0 & 0 & 1-2r+c_3
\end{array}
\right) \nonumber\\
&=&\frac{1}{4}
\left(
\begin{array}{cccc}
1+r+s+c_3
& 0 & 0 & c_1 -c_2 \\
0 & 1+r-s-c_3 & c_1+c_2 & 0 \\
0 & c_1 +c_2 & 1-r+s-c_3
& 0 \\
c_1 -c_2 & 0 & 0 & 1-r-s+c_3
\end{array}
\right) \,.
\end{eqnarray}
\end{widetext}

 The state in Eq. (\ref{state3}) has the following form
\begin{eqnarray}
\rho^{ab}=\frac{1}{4}(I\otimes I+r\sigma_{3}\otimes I+I\otimes s\sigma_{3}+\sum_{i=1}^3c_i\sigma_i\otimes\sigma_i),
\end{eqnarray}
where $\sigma_{1}, \sigma_{2}, \sigma_{3}$ is Pauli matrix.

The eigenvalues of the $X$ states in Eq. (\ref{state3}) are given by
\begin{eqnarray}
u_\pm=\frac{1}{4}(1-c_3\pm2|c| ),\nonumber\\
v_\pm=\frac{1}{4}(1+c_3\pm2|r|).\nonumber
\end{eqnarray}

The entropy is given by
\begin{eqnarray}\label{entropy3}
S(\rho)&=&2-\frac{1}{4}[(1-c_3+2|c| )\log(1-c_3+2|c| )\nonumber\\
       & &+(1-c_3-2|c| )\log(1-c_3-2|c| )\nonumber\\
       & &+(1+c_3+2|r|)\log(1+c_3+2|r|)\nonumber\\
       & &+(1+c_3-2|r|)\log(1+c_3-2|r|).\nonumber\\
\end{eqnarray}

Next, we evaluate the one-way deficit of the $X$ states in Eq. (\ref{state3}). Let $\{\Pi_{k}=|k\rangle\langle k|, k=0, 1\}$
be the local measurement for the party $b$ along the computational base ${|k\rangle}$;  then any von Neumann measurement for the party $b$ can be written as
\begin{eqnarray}\label{bk}
\{B_{k}=V\Pi_{k}V^{\dag}: k=0, 1\}
\end{eqnarray}
for some unitary $V\in U(2)$. For any unitary $V$,
\begin{eqnarray}\label{v}
V=tI+i\vec{y}\cdot\vec{\sigma}=\left(
                                 \begin{array}{cc}
                                   t+y_{3}i & y_{2}+y_{1}i \\
                                   -y_{2}+y_{1}i & t-y_{3}i \\
                                 \end{array}
                               \right).
\end{eqnarray}
with $t\in \mathbb{R}$, $\vec{y}=(y_{1}, y_{2}, y_{3})\in \mathbb{R}^{3}$, and
\begin{eqnarray}\label{ty}
t^{2}+y_{1}^{2}+y_{2}^{2}+y_{3}^{2}=1,
\end{eqnarray}
after the measurement ${B_{k}}$, the state $\rho^{ab}$ will be changed into the ensemble $\{{\rho_{k}, p_{k}}\}$ with
\begin{eqnarray}\label{pp}
\rho_{k}: =\frac{1}{p_{k}}(I\otimes B_{k})\rho(I\otimes B_{k}),~~p_{k}=tr(I\otimes B_{k})\rho(I\otimes B_{k}).
\end{eqnarray}
To evaluate $\rho_{k}$ and $p_{k}$, we write
\begin{eqnarray}
p_{k}\rho_{k}&=&(I\otimes B_{k})\rho(I\otimes B_{k})\nonumber\\
&=&\frac{1}{4}(I\otimes V)(I\otimes \Pi_{k})[I+r\sigma_{3}\otimes I+sI\otimes V^{\dag}\sigma_{3}V^{\dag}\nonumber\\
& &+\sum_{j=1}^{3} c_{j}\sigma_{j}\otimes (V^{\dag} \sigma_{j} V)](I\otimes \Pi_{k})(I\otimes V^{\dag}).\nonumber
\end{eqnarray}
By the relations \cite{luo}
\begin{eqnarray}
V^{\dag}\sigma_{1}V&=&(t^{2}+y_{1}^{2}-y_{2}^{2}-y_{3}^{2})\sigma_{1}+2(ty_{3}+y_{1}y_{2})\sigma_{2}\nonumber\\
& &+2(-ty_{2}+y_{1}y_{3})\sigma_{3},\label{condition2} \nonumber \\
V^{\dag}\sigma_{2}V&=&2(-ty_{3}+y_{1}y_{2})\sigma_{1}+(t^{2}+y_{2}^{2}-y_{1}^{2}-y_{3}^{2})\sigma_{2}\nonumber\\
& &+2(ty_{1}+y_{2}y_{3})\sigma_{3},\label{condition3} \nonumber\\
V^{\dag}\sigma_{3}V&=&2(ty_{2}+y_{1}y_{3})\sigma_{1}+2(-ty_{1}+y_{2}y_{3})\sigma_{2}\nonumber\\
& &+(t^{2}+y_{3}^{2}-y_{1}^{2}-y_{2}^{2})\sigma_{3},\label{condition4}\nonumber
\end{eqnarray}
and
\begin{eqnarray}\label{condition5}
\Pi_{0}\sigma_{3}\Pi_{0}=\Pi_{0}, \Pi_{1}\sigma_{3}\Pi_{1}=-\Pi_{1},\Pi_{j}\sigma_{k}\Pi_{j}=0,\nonumber\\
\end{eqnarray}
for $j=0, 1, k=1, 2$,
we obtain
\begin{eqnarray}
p_{0}\rho_{0}&=&\frac{1}{4}[I+rz_{3}I+cz_{1}\sigma_{1}+cz_{2}\sigma_{2}+(r+c_{3}z_{3})\sigma_{3}]\nonumber\\
& &\otimes(V\Pi_{0}V^{\dag}),\nonumber\\
p_{1}\rho_{1}&=&\frac{1}{4}[I-rz_{3}I-cz_{1}\sigma_{1}-cz_{2}\sigma_{2}+(r-c_{3}z_{3})\sigma_{3}]\nonumber\\
& &\otimes(V\Pi_{1}V^{\dag}),\nonumber
\end{eqnarray}
where
\begin{eqnarray}
z_{1}&=&2(-ty_{2}+y_{1}y_{3}),\nonumber\\
z_{2}&=&2(ty_{1}+y_{2}y_{3}), \nonumber\\
z_{3}&=&t^{2}+y_{3}^{2}-y_{1}^{2}-y_{2}^{2}.\label{condition6}\nonumber
\end{eqnarray}
Then, we will evaluate the eigenvalues of $\sum\limits_{k}\Pi_{k}\rho^{ab}\Pi_{k}$ by
\begin{eqnarray}
\sum\limits_{k}\Pi_{k}\rho^{ab}\Pi_{k}=p_{0}\rho_{0}+p_{1}\rho_{1},
\end{eqnarray}
and
\begin{eqnarray}
& &p_{0}\rho_{0}+p_{1}\rho_{1}\nonumber\\
&=&\frac{1}{4}(I+r\sigma_{3})\otimes I\nonumber\\
& &+\frac{1}{4}(rz_{3}I+cz_{1}\sigma_{1}+cz_{2}\sigma_{2}+c_{3}z_{3}\sigma_{3})
\otimes V\sigma_{3}V^{\dag}.\nonumber
\end{eqnarray}
The eigenvalues of $p_{0}\rho_{0}+p_{1}\rho_{1}$  are the same with the eigenvalues of the states $(I\otimes V^{\dag})(p_{0}\rho_{0}+p_{1}\rho_{1})(I\otimes V)$, and
\begin{eqnarray}\label{value2}
& &(I\otimes V^{\dag})(p_{0}\rho_{0}+p_{1}\rho_{1})(I\otimes V)\nonumber\\
&=&\frac{1}{4}(I+r\sigma_{3})\otimes I\nonumber\\
& &+\frac{1}{4}(rz_{3}I+cz_{1}\sigma_{1}+cz_{2}\sigma_{2}+c_{3}z_{3}\sigma_{3})\otimes\sigma_{3}.
\end{eqnarray}
By using  $z_{1}^{2}+z_{2}^{2}+z_{3}^{2}=1$, the eigenvalues of the states in the equation (\ref{value2}) are
\begin{eqnarray}
\omega_{1,2}
&=&\frac{1}{4}\left(1-rz_{3}\pm\sqrt{r^{2}-2rc_{3}z_{3}+c^{2}+(c_{3}^{2}-c^{2})z_{3}^{2}}\right),\nonumber
\end{eqnarray}
\begin{eqnarray}
\omega_{3,4}
&=&\frac{1}{4}\left(1+rz_{3}\pm\sqrt{r^{2}+2rc_{3}z_{3}+c^{2}+(c_{3}^{2}-c^{2})z_{3}^{2}}\right).\nonumber
\end{eqnarray}

The entropy of $\sum\limits_{k}\Pi_{k}\rho^{ab}\Pi_{k}$ is $S(\sum\limits_{k}\Pi_{k}\rho^{ab}\Pi_{k})=-\sum\limits_{i=1}^{4}\omega_{i}\log\omega_{i}$. When $\lambda$ is fixed, $r, c, c_{3}$ is constant.
It converts the problem about $\min \limits_{\{\Pi_{k}\}} S(\sum\limits_{k}\Pi_{k}\rho^{ab}\Pi_{k})$ to the problem about the function of one variable $z_{3}$ for minimum. That is
\begin{eqnarray}\label{min3}
\min \limits_{\{\Pi_{k}\}} S(\sum\limits_{k}\Pi_{k}\rho^{ab}\Pi_{k})=\min \limits_{\{z_{3}\}} S(\sum\limits_{k}\Pi_{k}\rho^{ab}\Pi_{k}).
\end{eqnarray}


By Eqs. (\ref{definition}), (\ref{entropy3}), (\ref{min3}), the one-way deficit of the $X$ states in Eq. (\ref{state3}) is given by
\begin{eqnarray}
& &\Delta^{\rightarrow}(\rho^{ab})\nonumber\\
&=&\min\limits_{\{\Pi_{k}\}}S(\sum\limits_{k}\Pi_{k}\rho^{ab}\Pi_{k})-S(\rho^{ab})\nonumber\\
&=&\min\limits_{z_{3}}\{-\frac{1}{4}[\left(1-rz_{3}+\sqrt{r^{2}-2rc_{3}z_{3}+c^{2}+(c_{3}^{2}-c^{2})z_{3}^{2}}\right)\nonumber\\
& &\cdot\log\left(1-rz_{3}+\sqrt{r^{2}-2rc_{3}z_{3}+c^{2}+(c_{3}^{2}-c^{2})z_{3}^{2}}\right)\nonumber\\
& &+\left(1-rz_{3}-\sqrt{r^{2}-2rc_{3}z_{3}+c^{2}+(c_{3}^{2}-c^{2})z_{3}^{2}}\right)\nonumber\\
& &\cdot\log\left(1-rz_{3}-\sqrt{r^{2}-2rc_{3}z_{3}+c^{2}+(c_{3}^{2}-c^{2})z_{3}^{2}}\right)\nonumber\\
& &+\left(1+rz_{3}+\sqrt{r^{2}+2rc_{3}z_{3}+c^{2}+(c_{3}^{2}-c^{2})z_{3}^{2}}\right)\nonumber\\
& &\cdot\log\left(1+rz_{3}+\sqrt{r^{2}+2rc_{3}z_{3}+c^{2}+(c_{3}^{2}-c^{2})z_{3}^{2}}\right)\nonumber\\
& &+\left(1+rz_{3}-\sqrt{r^{2}+2rc_{3}z_{3}+c^{2}+(c_{3}^{2}-c^{2})z_{3}^{2}}\right)\nonumber\\
& &\cdot\log\left(1+rz_{3}-\sqrt{r^{2}+2rc_{3}z_{3}+c^{2}+(c_{3}^{2}-c^{2})z_{3}^{2}}\right)]\}\nonumber\\
& &+\frac{1}{4}[(1-c_3+2|c| )\log(1-c_3+2|c| )\nonumber\\
       & &+(1-c_3-2|c| )\log(1-c_3-2|c| )\nonumber\\
       & &+(1+c_3+2|r|)\log(1+c_3+2|r|)\nonumber\\
       & &+(1+c_3-2|r|)\log(1+c_3-2|r|)].\nonumber\\
\end{eqnarray}
Note that  $\Delta^{\rightarrow}(\rho^{ab})$ is an even function for the variable $z_{3}$, so we can focus on $z_{3}\in[0,1]$
instead of $[-1,1]$.

For example, we set $\lambda=0.6$, then $r=0.409666, c=-0.509296, c_{3}=-0.0915564$,
and obtain that the value of the one-way deficit is $0.418314$.

\section{One-way deficit in $XX$ model for $\lambda>1$}

When $\lambda>1$, $\Lambda_k$ is negative for all $k$. Therefore the ground state has $N$ fermions, which is translated as all spins up in spin language. It is straightforward to see that in this case the one-way deficit of the ground state or that of any part of its reduced density operator is always zero.

In fact, for $\lambda >1$ the ground state is polarized, whose reduced density matrix for two spins at $l$ and $l+1$ has turned into
\begin{eqnarray}\label{state2}
\rho_{l,l+1}=|\uparrow \uparrow\rangle \langle\uparrow \uparrow|
\end{eqnarray}
The density matrix in Eq. (\ref{state2}) is rewritten as
\begin{eqnarray}\label{state4}
\rho_{l,l+1}=|0\rangle\langle0|\otimes|0\rangle\langle0|
\end{eqnarray}

Next, we evaluate the one-way deficit of the states in Eq. (\ref{state4}). Let $\{\Pi_{k}=|k\rangle\langle k|, k=0, 1\}$
be the local measurement for the party $b$ along the computational base ${|k\rangle}$. By (\ref{bk}), (\ref{v}), (\ref{pp}), after the measurement ${B_{k}}$, the state in Eq. (\ref{state4}) will be changed into the ensemble $\{{\bar{\rho}_{k}, \bar{p}_{k}}\}$. After some algebraic calculus, we obtain
\begin{eqnarray}
\bar{p}_{0}\bar{\rho}_{0}=(t^{2}+y_{3}^{2})|0\rangle\langle0|\otimes V|0\rangle\langle0|V^{\dag},\nonumber\\
\bar{p}_{1}\bar{\rho}_{1}=(y_{1}^{2}+y_{2}^{2})|0\rangle\langle0|\otimes V|1\rangle\langle1|V^{\dag}\nonumber
\end{eqnarray}
Then, we will evaluate the eigenvalues of $\sum\limits_{k}\Pi_{k}\rho_{l,l+1}\Pi_{k}$ by
\begin{eqnarray}
\sum\limits_{k}\Pi_{k}\rho_{l,l+1}\Pi_{k}=\bar{p}_{0}\bar{\rho}_{0}+\bar{p}_{1}\bar{\rho}_{1},
\end{eqnarray}
and
\begin{eqnarray}
\bar{p}_{0}\bar{\rho}_{0}+\bar{p}_{1}\bar{\rho}_{1}=|0\rangle\langle0|\otimes V\left(
                                                                                 \begin{array}{cc}
                                                                                   t^{2}+y_{3}^{2} & 0 \\
                                                                                   0 & y_{1}^{2}+y_{2}^{2} \\
                                                                                 \end{array}
                                                                               \right)V^{\dag}.
\end{eqnarray}
The eigenvalues of $\bar{p}_{0}\bar{\rho}_{0}+\bar{p}_{1}\bar{\rho}_{1}$  are
$t^{2}+y_{3}^{2}, y_{1}^{2}+y_{2}^{2}, 0, 0$. The entropy of $\sum\limits_{k}\Pi_{k}\rho_{l,l+1}\Pi_{k}$ is
\begin{eqnarray}
& &S(\sum\limits_{k}\Pi_{k}\rho_{l,l+1}\Pi_{k})\nonumber\\
&=&-(t^{2}+y_{3}^{2})\log(t^{2}+y_{3}^{2})\nonumber\\
& &-(y_{1}^{2}+y_{2}^{2})\log(y_{1}^{2}+y_{2}^{2}).
\end{eqnarray}
By Eq. (\ref{ty}), we know that $S(\sum\limits_{k}\Pi_{k}\rho_{l,l+1}\Pi_{k})$ is binary entropy function. When $t=0, y_{3}=0, y_{1}^{2}+y_{2}^{2}=1$ or $y_{1}=0, y_{2}=0, t^{2}+y_{3}^{2}=1$, the function $S(\sum\limits_{k}\Pi_{k}\rho_{l,l+1}\Pi_{k})$ reaches the minimum $0$. By the entropy of $\rho_{l,l+1}$ in Eq. (\ref{state4}) being zero and Eq. (\ref{definition}), the one-way deficit of the states in Eq. (\ref{state4}) is $0$.

In Fig. 1, we drawn the curve of one-way deficit of two adjacent spins in the bulk of $XX$
model in $\lambda\in[0,1.5]$.

We find that the one-way deficit
is nonzero in the domain $\lambda\in[0,1)$ and then becomes zero when $\lambda\geq1$.
As the $XX$ model undergoes a first order
transition at the critical point $\lambda=1$ from fully polarized to a
critical phase with quasi-long-range order,  we conclude that one-way deficit can be used to
detect quantum phase of the $XX$ model, and moreover, may reveal the insight of phase
transition by quantum correlations.

\begin{figure}[ht]
\includegraphics[scale=0.5]{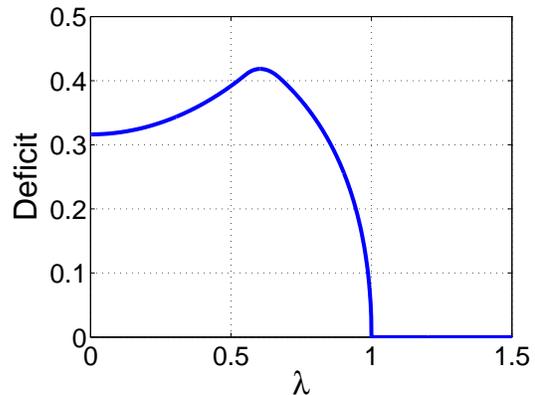}
\caption{(Color online) One-way deficit of two adjacent spins in the bulk for the $XX$ model in the
thermodynamic limit as a function of the quantum parameter $\lambda$.}
\end{figure}

\section{conclusion}\label{sec:conclusion}

We have given a method to evaluate the one-way deficit two adjacent spins in the bulk for the $XX$ model in the thermodynamic limit. We have drawn the curve of one-way deficit of the $XX$ model. We find that we can use one-way deficit to detect quantum phase of the $XX$ model.
We find that the one-way deficit becomes zero when $\lambda\geq1$. Therefore, the one-way deficit can characterizes
the quantum phase transition in the XX model.
This may shed lights on the study of properties of quantum correlations
in different quantum phases.

\begin{acknowledgments}
We are thankful to Yu-Ran Zhang and Jin-Ju Chen for fruitful discussions. This work was supported by the Science and Technology Research Plan Project of the Department of Education of Jilin Province in the Twelfth Five-Year Plan, the National Natural Science Foundation of China under grant Nos. 11175248, 11275131, 11305105.
\end{acknowledgments}

\end{document}